\documentclass[preprint,aps,12pt,showpacs,nofootinbib,tightenlines]{revtex4}
\usepackage{amsmath}
\usepackage{amssymb}
\usepackage{epsfig}
\usepackage{graphicx}
\begin{document}
\preprint{NJNU-TH-06-30}
\newcommand{\be}{\begin{equation}}
\newcommand{\ee}{\end{equation}}
\newcommand{\beq}{\begin{eqnarray}}
\newcommand{\eeq}{\end{eqnarray}}

\newcommand{\ov}{\overline }
\newcommand{\nnb}{\nonumber}
\newcommand{\non}{\nonumber\\ }
\def \epjc{  Eur. Phys. J. C }
\def \jpg{  J. Phys. G }
\def \jhep{  J. High Energy Phys. }
\def \npb{  Nucl. Phys. B }
\def \plb{  Phys. Lett. B }
\def \prd{  Phys. Rev. D }
\def \prl{  Phys. Rev. Lett.  }
\def \rmp{  Rev. Mod. Phys. }

\title{ $B^0_{s(d)} - \bar{B}^0_{s(d)} $ mixing and new physics effects in a top quark
two-Higgs doublet model}
\author{Lin-xia L\"u $^{a,b}$} \email{lulinxia@email.njnu.edu.cn}
\author{Zhen-jun Xiao $^a$} \email{xiaozhenjun@njnu.edu.cn}
\affiliation{a.  Department of Physics and Institute of Theoretical Physics, Nanjing Normal
University, Nanjing, Jiangsu 210097, P.R.China}
\affiliation{b.  Department of Physics, Nanyang Teacher's College, Nanyang,
Henan 473061, P.R.China}
\date{\today}
\begin{abstract}
We calculate the new physics contributions to the neutral $B_d^0$ and $B_s^0$ meson
mass splitting $\Delta M_d$ and $\Delta M_s$ induced by the box diagrams involving the charged-Higgs bosons
in the top quark two-Higgs doublet model (T2HDM).
Using the precision data, we obtain the bounds on the parameter space of the T2HDM:
(a) for fixed $M_H=400$ GeV and $\delta=[0^\circ,60^\circ]$, the upper bound on $\tan{\beta}$ is
$\tan \beta \leq 30$ after the inclusion of major theoretical uncertainties;
(b) for the case of $\tan{\beta} \leq 20$, a light charged Higgs boson with a mass around
$300$ GeV is allowed; and (c) the bounds on $\tan{\beta}$ and $M_H$ are strongly correlated:
a smaller (larger) $\tan{\beta}$ means a lighter (heavier) charged Higgs boson.
\end{abstract}


\maketitle


\newpage
\section{introduction}\label{sec-1}

As a flavor changing neutral current  process, $B_q^0 -
\bar{B}_q^0$ mixing with $q=d,s$ are generated at the loop-level and
have been of fundamental importance in probing virtual effects from potential new
physics beyond the standard model (SM). The $B_q^0 - \bar{B}_q^0$ mixing is
responsible for the small mass differences between the heavy and light mass
eigenstates of neutral B mesons:
\beq
\Delta M_q = M_{B_{q,H}^0} - M_{B_{q,L}^0}. \label{eq:dmq1}
\eeq
The mass splitting $\Delta M_d$ has been measured with high precision \cite{pdg06,hfag06},
while the measurement of $\Delta M_s$ is very difficult due to the rapid oscillation of $B_s$ meson
and has been reported  by CDF and D$0$ Collaboration \cite{cdf06a,cdf06b,d006} very recently.
The world average for $\Delta M_d$ \cite{hfag06} and the first observation of $\Delta M_s$
from CDF \cite{cdf06b} are the following
\beq
\Delta M_d &=& 0.507 \pm 0.005\; {\rm  ps^{-1}}, \label{eq:exp1} \\
\Delta M_s &=& 17.77 \pm 0.10({\rm stat})\pm 0.07 ({\rm syst}) {\rm ps^{-1}},
\label{eq:exp2}
\eeq
which agree well with the standard model (SM) predictions or the results from global fit
~\cite{utfit}. The perfect agreement between the SM prediction and the experimental measurements
permit us to put strong constraints on the parameter spaces of various new physics models.

In the SM, $B^0-\bar{B}^0$ mixing is dominated by the box diagrams
with two internal t-quarks and W gauge bosons. In new physics
models, the box diagrams with one or two W gauge bosons replaced by
the new charged scalars or vector bosons and/or top quarks replaced
by new fermions can also contribute to $B^0 - \bar{B}^0$ mixing.
Using the precision data, we study both $B_d^0 - \bar{B}_d^0$ and $B_s^0 - \bar{B}_s^0$
mixing in the top quark two-Higgs doublet model (T2HDM) and try to find the constrains on
the parameter space of this model.

During the past years, $B^0 - \bar{B}^0$ mixing has been studied extensively
in the SM and various new physics models. The charged-Higgs
boson contributions to $B^0 - \bar{B}^0$ mixing have been calculated
at the leading order (LO) for a long time~\cite{asw80}. The next-to-leading order (NLO)
quantum chromodynamics (QCD) correction to $B^0 - \bar{B}^0$ mixing
is firstly presented in Ref.~\cite{buras90} and the analytic formulae for the QCD
renormalization group factors are given in Ref.~\cite{buras605}.
In Ref.~\cite{urban98}, the authors studied the new physics effects in
the conventional two-Higgs-doublet model (2HDM) of type I and II.
In Refs.~\cite{huang2004,grant95,xiao04} the authors have
calculated the charged Higgs boson contributions to the mass splitting
$\Delta M_{B_d}$ and drawn the constraints on the parameters of the third type of
2HDM (the model III) at the LO or NLO level.
Very recently, $B_s^0 - \bar{B}_s^0$ mixing has been used to put
constraints on various new physics models, for example, in Refs.~\cite{newp}
after the release of the new date of $\Delta M_s$.
In this paper, we will calculate the new physics contributions to $\Delta M_q$ within
the framework of T2HDM \cite{das1996,kiers,kiers62}. By comparing the theoretical predictions
with the precision data, we draw the constraints on the free
parameters of T2HDM.

The organization of this paper is as follows. In sec.~\ref{sec-2}, we firstly give
a brief review for the top quark two-Higgs doublet model, and then present
the one-loop contributions to the mass splitting  $\Delta M_d$ and $\Delta M_s$ induced by
the box diagrams involving the charged Higgs bosons in the T2HDM. Numerical results
are presented in sec.~\ref{sec-3}, and the conclusions are included in the
final section.


\section{$B^0 - \bar{B}^0$ mixing in the T2HDM}\label{sec-2}

The new physics  model considered here is the T2HDM proposed in Ref.~\cite{das1996} and
studied for example in Refs.~\cite{kiers,kiers62,xiao06},
which is also a special case of the 2HDM of type III~\cite{hou1992}.
This model is designed to accommodate the heaviness of the top quark by coupling it to a
scalar doublet with large vacuum expectation value (VEV). All the
other five quarks are coupled to another scalar doublet, whose VEV is
much smaller. As a result, $\tan\beta$ is naturally large in this model.

Let us now briefly recapitulate some important features of the T2HDM \cite{das1996}.
Consider the Yukawa Lagrangian of the form:
\beq {\cal L }_Y = - {\overline{L}}_L \phi_1 E l_R -
{\overline{Q}}_L \phi_1 F d_R - {\overline{Q}}_L
{\widetilde{\phi}}_1 G {\bf 1}^{(1)} u_R -
{\overline{Q}}_L{\widetilde{\phi}}_2 G {\bf 1}^{(2)} u_R + H.c. \eeq
where $Q_L$ and $L_L$ are 3-vector of the left-handed quark and
lepton doublets; $\phi_i$ $(i=1,2)$ are the two Higgs doublets
with ${\widetilde{\phi}}_i = i \tau_2 \phi^*_i $; and $E$, $F$ and $G$
are the $3 \times 3$ matrices in the generation space and give
masses respectively to the charged leptons, the down and up type
quarks; ${\bf 1} ^{(1)} \equiv diag(1,1,0)$ and ${\bf 1} ^{(2)}
\equiv diag(0,0,1)$ are the two orthogonal projection operators
onto the first two and the third families respectively.

The Yukawa couplings involving the charged-Higgs bosons are of the
form \cite{das1996}
\beq \label{eq:Lagrangian} {\cal L}^C_Y  &=&\frac{g}{\sqrt{2}M_W} \{
- \overline{u}_L V M_D d_R [G^+ - \tan \beta H^+ ] +\overline{u}_R
M_U V d_L [G^+ - \tan \beta H^+] \non
 &&  + \overline{u}_R \Sigma^ {\dag} V d_L [\tan \beta + \cot \beta] H^+ +h.c. \}.
\eeq
where $G^{\pm}$ and $H^{\pm}$ denote the would-be Goldstone bosons
and the physical charged Higgs bosons, respectively. Here $M_U$ and
$M_D$ are the diagonal up- and down-type mass matrices, $V$ is the
usual CKM matrix and $\Sigma \equiv M_U U^{\dag}_R {\bf 1}^{(2)}
U_R$. $U^{\dag}_R$ is the unitary matrix which diagonalizes the
right-handed up-type quarks and has the following form:
\beq \label{eq: UR}
U_R = \left(\begin{array}{ccc} \cos\phi & -\sin\phi & 0 \\
\sin\phi & \cos\phi & 0 \\
 0 & 0 & 1 \end{array} \right) \times \left(\begin{array}{ccc}
1 & 0 & 0 \\  0 & \sqrt{1-|\epsilon_{ct} \xi|^2} & -\epsilon_{ct}
\xi^* \\ 0 & \epsilon_{ct} \xi & \sqrt{1-|\epsilon_{ct} \xi|^2}
\end{array} \right).  \eeq
where $\epsilon_{ct} \equiv m_c/m_t$, and $\xi=|\xi|e^{i\delta }$ is
a complex number of order unity. Inserting Eq.~(\ref{eq: UR}) into
the definition of $\Sigma$ yields
\beq \label{eq:sigma}
\Sigma= \left(\begin{array}{ccc} 0 & 0 & 0 \\
0 & m_c \epsilon_{ct}^2 |\xi|^2 & m_c \epsilon_{ct}
\xi^*\sqrt{1-|\epsilon_{ct} \xi|^2} \\ 0 & m_c \xi
\sqrt{1-|\epsilon_{ct} \xi|^2} & m_t (1-|\epsilon_{ct} \xi|^2)
\end{array} \right).
\eeq

Now we are ready to calculate the charged Higgs and
would-be Goldstone bosons contributions to the mass splitting
$\Delta M_{B_q} (q=d,s)$ in the T2HDM.


The effective weak Hamiltonian for $\Delta B = 2$ processes beyond
the SM can be written as
\begin{eqnarray}
{\cal H}_{\rm eff}^{\Delta B=2} &=& \sum_{i=1}^{5} C_i Q_i +
  \sum_{i=1}^{3} \tilde{C}_i \tilde{Q}_i
\label{eq:heff}
\end{eqnarray}
with
\beq
Q_1 &=& \left ( \bar{q}^{\alpha}_L \gamma^\mu b^{\alpha}_L \right ) \left ( \bar{q}^{\beta}_{L}
  \gamma_\mu b^{\beta}_L \right ),  \non
Q_2 &=& \left ( \bar{q}^{\alpha}_R  b^{\alpha}_L \right )  \left (\bar{q}^{\beta}_R b^{\beta}_L\right ),
\qquad
Q_3 = \left ( \bar{q}^{\alpha}_R  b^{\beta}_L  \right ) \left ( \bar{q}^{\beta}_R b^{\alpha}_L\right ),
\non
Q_4 &=& \left ( \bar{q}^{\alpha}_R  b^{\alpha}_L \right ) \left (\bar{q}^{\beta}_L b^{\beta}_R\right ),
\qquad
Q_5 = \left (\bar{q}^{\alpha}_R  b^{\beta}_L \right )  \left (\bar{q}^{\beta}_L b^{\alpha}_R \right ),
\label{eq:ops}
\eeq
where $q = s,\,d$, corresponding to the operators of $B_s$ and $B_d$
system respectively, $P_{L,R} \equiv (1 \mp \gamma_5)/2$, and
$\alpha$, $\beta$ are color indexes. The tilde operators
$\tilde{Q}_i$ ($i=1,2,3$) correspond to the ones $Q_i$ ($i=1,2,3$)
with opposite chirality.

In T2HDM, there are two CP-even scalars ($H^0$, $h^0$), one CP-odd scalar ($A^0$), two
charged Higgs bosons ($H^{\pm}$), and the Goldstone bosons ($G^{\pm}$, $G^0$).
At one loop level only the charged scalars are relevant for the box
diagrams contributing to the $B^0 - \bar{B}^0$ mixing amplitude.
From Eq.~(\ref{eq:Lagrangian}), we can rewrite the vertex couplings
of $H_l^+ \equiv ( H^+, G^+)$~\cite{notation} (where $G^\pm$ is the
would-be Goldstone boson) in a compact form
\beq
{\cal{L}}_{int} = H^+_l \bar{u}_A V_{AI} \left (a_L^{AIl} P_L + a_R^{AIl}
P_R \right ) d_I + h.c.
\eeq
where
\beq
a_L^{AIl} &=& \frac{e}{\sqrt{2} s_W} \frac{m_A }{M_W} \; \cdot \;
\left\{ \begin{array}{ll}
\,\, \left(\frac{(\Sigma^{\dag}V)_{AI}}{m_A V_{AI}} -1\right)\tan\beta,
\hspace{1cm} & \textrm{ for \, l=1,} \\
\,\,\,\,\, 1, \hspace{2cm} & \textrm{ for \, l=2,}
\end{array} \right. \\
a_R^{AIl} &=& \frac{e}{\sqrt{2} s_W} \frac{m_I }{M_W} \; \cdot \;
\left\{ \begin{array}{ll}
\,\,\tan\beta , \hspace{1cm} & \textrm{ for \, l=1 ,} \\
\,\,\,\,\, -1 , \hspace{2cm} & \textrm{ for \, l=2 ,}
\end{array} \right.
\eeq
with $m_A=(m_u, m_c, m_t)$ and $m_I=(m_d,m_s,m_b)$. The contributions of $H_l^+$ to the
Wilson coefficients $C_i$ of the relevant operators responsible for
$B^0 - \bar{B}^0$ mixing can be easily expressed in terms of the
coefficients $a_L^{AIl} $ and $a_R^{AIl} $as follows.

The contributions to the Wilson $C_1(\mu)$ and $C_2(\mu)$ induced by the box diagrams
with one $W^\pm$ and one $H^\pm$ propagator can be written as
\footnote{The contribution of $G^\pm$ to $C_1(\mu)$ is already taken into account
in the Inami-Lim function $S_0(x_t)$~\cite{Inami-Lim}. Mass of the
$u$ quark are neglected.}:
\begin{eqnarray}
C_1(\mu)&=&\frac{e^2}{2 s^2_W} \sum_{A,A'} \frac{V_{AI}^\ast
V_{AJ}V_{A'I}^\ast V_{A'J}}{16 \pi^2} m_A m_{A'}  a_L^{\dagger
AI1} a_L^{A'J1} \cdot D_0(m_A^2, m_{A'}^2,M_W^2,M_{H^+}^2 ), \non
C_2(\mu)&=&{e^2\over2s^2_W}\sum_{A,A'}\sum_{l=1}^2\frac{V_{AI}^\ast
V_{AJ}V_{A'I}^\ast V_{A'J}}{16 \pi^2} a_R^{\dagger AIl} a_R^{A'Jl}\cdot 4
D_{00}(m_A^2,m_{A'}^2,M_W^2,m_{H_l^+}^2),
\end{eqnarray}
where $s_W \equiv \sin \theta_W$ ($\theta_W$ is the Weinberg angle), while
the four-point integral functions $D_0$ and $D_{00}$ can be written as
\beq
D_0(a,b,c,d)&=&\int \frac{d^4q}{i \pi^2}\frac{1}{(q^2-a)(q^2-b)(q^2-c)(q^2-d)}\non &=& \frac{a
}{(b-a)(c-a)(d-a)}
\log[\frac{a}{d}]+\frac{b}{(a-b)(b-c)(b-d)}\log[\frac{b}{d}]  \non &
& +\frac{c}{(a-c)(b-c)(d-c)}\log[\frac{c}{d}] \; , \\
D_{00}(a,b,c,d)&=&\frac{1}{4}\int \frac{d^4q}{i
\pi^2}\frac{q^2}{(q^2-a)(q^2-b)(q^2-c)(q^2-d)}\non &=&
\frac{a^2}{(b-a)(c-a)(d-a)}
\log[\frac{a}{d}]+\frac{b^2}{(a-b)(b-c)(b-d)}\log[\frac{b}{d}]  \non
& & +\frac{c^2}{(a-c)(b-c)(d-c)}\log[\frac{c}{d}] \; .
\eeq

\begin{table}[]
\begin{center}
\caption{The "magic numbers" appearing in the calculation of the
Wilson coefficients in the process of $B^0 - \ov{B}^0$ mixing.}
\label{tab:magic}
\vspace{0.2cm}
\begin{tabular}{ c|ccccc|c|ccccc } \hline\hline
 & 1 & 2 & 3 & 4 & 5 &  & 1 & 2 & 3 & 4 & 5  \\ \hline
$a_i$  &0.286& -0.692& 0.787 &-1.143 & 0.143 & & & & & &   \\ \hline
$b_i^{(11)}$ &0.865 & 0 & 0 & 0 & 0 &
$c_i^{(11)}$ &-0.017 & 0&0 &0 &0  \\
$b_i^{(22)}$ &0 &1.879&0.012&  0 & 0 &
$c_i^{(22)}$ &0 & -0.18&-0.003&0 &0 \\
$b_i^{(23)}$ &0 &-0.493 & 0.18 & 0 & 0 &
$c_i^{(23)}$ &0 &-0.014&0.008 &0 &0 \\
$b_i^{(32)}$ &0 &-0.044 & 0.035 & 0 & 0 &
$c_i^{(32)}$ &0 &0.005&-0.012 &0 &0 \\
$b_i^{(33)}$ &0 &0.011&0.54 & 0 & 0 &
$c_i^{(33)}$ &0 & 0.000 & 0.028&0 &0 \\
$b_i^{(44)}$ &0 &0 & 0& 2.87 & 0 &
$c_i^{(44)}$ &0 &0&0 &-0.48 &0.005 \\
$b_i^{(45)}$ &0 &0 & 0 & 0.961 & -0.22 &
$c_i^{(45)}$ &0 &0&0 &-0.25 &-0.006 \\
$b_i^{(54)}$ &0 &0 & 0 & 0.09 & 0 &
$c_i^{(54)}$ &0 & 0 & 0 &-0.013 &-0.016 \\
$b_i^{(55)}$ &0 &0 & 0 & 0.029 & 0.863 &
$c_i^{(55)}$ &0 &0&0 &-0.007 &0.019 \\
\hline \hline
\end{tabular}
\end{center}
\end{table}

The contributions induced by the box diagrams
with two $H_l^\pm$ propagators can be written as \footnote{In the sum over $l$ and
$n$ in the expression for $C_1(\mu)$ the contribution of $G^\pm
G^\mp$ is excluded since it has been taken into account in the
function $S_0(x_t)$~\cite{Inami-Lim}.}:
\beq
C_1(\mu)&=&-{1\over2}\sum_{A,A'}\sum_{l,n}\frac{V_{AI}^\ast
V_{AJ}V_{A'I}^\ast V_{A'J}}{16 \pi^2} a_L^{\dagger AIl} a_L^{AJn}
a_L^{\dagger A'In} a_L^{A'Jl} \non
&& \ \ \ \cdot D_{00}\left (m_A^2,m_{A'}^2,M_{H^+_l}^2,M_{H^+_n}^2 \right ),
\eeq
\beq
\tilde{C}_1(\mu) &=& -{1\over2}\sum_{A,A'} \sum_{l,n}^2
\frac{V_{AI}^\ast V_{AJ}V_{A'I}^\ast V_{A'J}}{16 \pi^2} a_R^{\dagger
AIl} a_R^{AJn} a_R^{\dagger A'In} a_R^{A'Jl} \non
&& \ \ \ \cdot D_{00}\left (m_{A}^2,m_{A'}^2,M_{H^+_l}^2,M_{H^+_n}^2\right ), \\
C_2(\mu) &=& -{1\over2}\sum_{A,A'} \sum_{l,n}^2 \frac{V_{AI}^\ast
V_{AJ}V_{A'I}^\ast V_{A'J}}{16 \pi^2}m_{u_A}m_{u_A'} a_R^{\dagger
AIl} a_L^{AJn} a_R^{\dagger A'In} a_L^{A'Jl}\non
&& \ \ \ \cdot D_0\left(m_{A}^2, m_{A'}^2, M_{H^+_l}^2, M_{H^+_n}^2 \right ), \\
\tilde{C}_2(\mu) &=& -{1\over2}\sum_{A,A'} \sum_{l,n}^2
\frac{V_{AI}^\ast V_{AJ}V_{A'I}^\ast V_{A'J}}{16
\pi^2}m_{u_A}m_{u_A'} a_L^{\dagger AIl} a_R^{AJn} a_L^{\dagger A'In} a_R^{A'Jl}\non
&& \ \ \ \cdot D_0\left (m_{A}^2, m_{A'}^2, M_{H^+_l}^2, M_{H^+_n}^2\right ), \\
C_4(\mu) &=& -\sum_{A,A'}\sum_{l,n}^2 \frac{V_{AI}^\ast
V_{AJ}V_{A'I}^\ast V_{A'J}}{16 \pi^2}m_{u_A}m_{u_A'} a_R^{\dagger
AIl} a_L^{AJn} a_L^{\dagger A'In} a_R^{A'Jl}\non
&& \ \ \ \cdot D_0\left (m_{A}^2, m_{A'}^2, M_{H^+_l}^2, M_{H^+_n}^2 \right ), \\
C_5(\mu) &=& 2\sum_{A,A'}\sum_{l,n}^2 \frac{V_{AI}^\ast
V_{AJ}V_{A'I}^\ast V_{A'J}}{16 \pi^2}a_L^{\dagger AIl} a_L^{AJn}
a_R^{\dagger A'In} a_R^{A'Jl}\non
&& \ \ \ \cdot D_{00}\left (m_{A}^2,m_{A'}^2,M_{H^+_l}^2,M_{H^+_n}^2\right ).
\label{eq:wilson}
\end{eqnarray}
These new physics contributions to the Wilson Coefficients are
consistent with the ones as given in Refs.~\cite{buras01}, the different sign
and factor are due to the different definitions of the four-point
integral functions. At one loop level there are no contributions to
the Wilson coefficients of the operators $Q_3$ and $\tilde{Q}_3$.

Now, one needs to run the Wilson coefficients from the scale of new physics $\mu_t \sim M_W$
down to the low energy scale $\mu_b \sim m_b$ by using the QCD renormalization group equations.
For the evolution of these coefficients, we follow Ref.~\cite{renorm},
\begin{equation}
\label{eq:evolution}
C_r(m_b )=\sum_{i=1}^{5} \sum_{s=1}^{5}
\left( b^{(r,s)}_i + \eta \,c^{(r,s)}_i\right) \eta^{a_i} \,C_s(\mu_t)
\end{equation}
where we have set the new physics scale $\mu_t = m_t$ and $\eta=\alpha_s(\mu_b)/\alpha_s(m_t)$.
The magic numbers $a_i$, $b_i^{(r,s)}$ and $c_i^{(r,s)}$ in Eq.~(\ref{eq:evolution}) are
listed in Table \ref{tab:magic}.

The off-diagonal element $M_{12}$ in the $2 \times 2$ effective
Hamiltonian causes the $B^0 - \bar{B}^0$ mixing. The mass difference
between the two mass eigenstates $\Delta M_q$ is described by
\beq
\Delta M_q=2|M_{12}^{(q)}|
\label{eq:m12a}
\eeq
with
\beq
M_{12}^{(q)}=\langle \bar{B}^0_q | {\cal{H}}_{eff}(\Delta
B=2)|B^0_q \rangle.
\label{eq:m12b}
\eeq

In the SM, the mass splitting $\Delta M_q$ is calculated from the box diagrams
of $B_q^0 - \bar{B}_q^0$ mixing, dominated
by t-quark exchange. At the NLO level, one fins that ~\cite{buras96}
\beq
\Delta M_q  = \frac{G_F^2 M_W^2}{6 \pi^2}  m_{B_q}
\left (\hat{B}_{B_q} f_{B_q}^2 \right ) \eta_B S_0(x_t) |V_{tq}V_{tb}^\ast|^2
\eeq
where $G_F$ is the Fermi constant, $M_W$ the mass of the $W$ boson,
$S_0(x_t)$ the Inami-Lim function~\cite{Inami-Lim} with
$x_t=m_t^2/M_W^2$. The NLO short-distance QCD correction gives
$\eta_B=0.552$, which is same for both $B^0_d$ and $B^0_s$ systems.
The non-perturbative quantities $\hat{B}_{B_q}$ are the bag
parameter and $f_{B_q}$ is the $B_q^0$ meson decay constant~\cite{buras96}.

In terms of the bag-parameters, the matrix elements of the operators
$Q_i$ and $\tilde{Q}_i$ are written as follows~\cite{renorm}:
\begin{eqnarray}
\label{eq:bpars}
\langle \bar B_q \vert  \widehat{Q}_{1} (\mu) \vert
B_q \rangle & = & \frac{1}{3} m_{B_q} f_{B_q}^{2} B_1^{(q)}(\mu),
\nonumber \\
\langle \bar B_q \vert \widehat{Q}_{2} (\mu) \vert B_q \rangle &=&
-\frac{5}{24} \left( \frac{ m_{B_q} }{ m_{b}(\mu) + m_q(\mu)
}\right)^{2} m_{B_q} f_{B_q}^{2}  B_2^{(q)}(\mu),
\nonumber  \\
\langle \bar B_q \vert \widehat{Q}_{3} (\mu) \vert B_q \rangle &=&
\frac{1}{24} \left( \frac{ m_{B_q} }{ m_{b}(\mu) + m_q(\mu)
}\right)^{2}
m_{B_q} f_{B_q}^{2}  B_{3}^{(q)}(\mu),
\nonumber\\
\langle \bar B_q \vert \widehat{Q}_{4} (\mu) \vert B_q\rangle &=&
\frac{1}{4} \left( \frac{ m_{B_q} }{ m_{b}(\mu) + m_q(\mu)
}\right)^{2}
m_{B_q} f_{B_q}^{2} B_{4}^{(q)}(\mu),
\nonumber\\
\langle \bar B_q \vert \widehat{Q}_{5} (\mu) \vert B_q \rangle &=&
\frac{1}{12} \left( \frac{m_{B_q}  }{ m_{b}(\mu) + m_q(\mu)
}\right)^{2} m_{B_q} f_{B_q}^{2}  B_{5}^{(q)}(\mu),
\end{eqnarray}
where $\widehat{Q}_{i}(\mu)$ are the operators renormalised
at the scale $\mu$, $B_i$ is the so-called bag factor. The matrix elements of
$\tilde{Q}_{1-3}$ are the same as that of $Q_{1-3}$.
We use the same definition of $B$ parameters $B_i^{(q)}(\mu)$ as in Ref.~\cite{bagpara}
and find numerically that
\begin{equation}
\begin{tabular}{ll}
$B_{1}^{(d)}(m_b) =0.87(4)^{+5}_{-4}$, \;\;\;
    & $B_{1}^{(s)}(m_b) =0.86(2)^{+5}_{-4}$,  \\
$B_{2}^{(d)}(m_b) = 0.82(3)(4)$, \;\;\;
    & $B_{2}^{(s)}(m_b) = 0.83(2)(4)$, \\
$B_{3}^{(d)}(m_b)=1.02(6)(9)$, \;\;\;
    & $B_{3}^{(s)}(m_b) = 1.03(4)(9)$, \\
$B_{4}^{(d)}(m_b) = 1.16(3)^{+5}_{-7}$, \;\;\;
    & $B_{4}^{(s)}(m_b) = 1.17(2)^{+5}_{-7}$, \\
$B_{5}^{(d)}(m_b) = 1.91(4)^{+22}_{-7}$, \;\;\;
    & $B_{5}^{(s)}(m_b) = 1.94(3)^{+23}_{-7}$.
\end{tabular}
\end{equation}
%

\section{Numerical Analysis}\label{sec-3}

\begin{table}[]
\begin{center}
\caption{The $\Delta M_q (q=d,s)$ in the SM and the T2HDM for $M_H=300$,
 $\tan{\beta}=10,30,50$ and $\delta=0^\circ$ (a), $30^\circ$ (b) and $60^\circ$ (c)
 within 1$\sigma$ range of the input hadronic parameters of JLQCD.  }
\label{tab:deltam} \vspace{0.2cm}
\begin{tabular}{ c|c|c|c|c} \hline\hline
& SM & \multicolumn{3}{|c}{T2HDM} \\
\cline{3-3}\cline{4-4}\cline{5-5}
& &$\tan{\beta}=10$&$\tan{\beta}=30$ &$\tan{\beta}=50$\\ \hline
                &                          &(a)$0.511^{+0.094}_{-0.133}$&(a)$0.634^{+0.115}_{-0.162}$ &(a)$1.630^{+0.289}_{-0.394}$ \\
$\Delta M_d$&$0.510^{+0.093}_{-0.133}$ &(b)$0.511^{+0.095}_{-0.132}$&(b)$0.665^{+0.121}_{-0.169}$ &(b)$1.737^{+0.309}_{-0.421}$ \\
                &                          &(c)$0.510^{+0.094}_{-0.132}$&(c)$0.569^{+0.115}_{-0.145}$ &(c)$1.472^{+0.260}_{-0.353}$ \\ \hline
                &                          &(a)$17.15^{+3.07}_{-2.69}$&(a)$20.09^{+3.55}_{-3.11}$ &(a)$44.76^{+7.61}_{-6.64}$ \\
$\Delta M_s$&$17.20^{+3.08}_{-2.69}$&(b)$17.12^{+3.06}_{-2.68}$&(b)$18.36^{+3.26}_{-2.86}$ &(b)$39.32^{+6.65}_{-5.80}$ \\
                &                          &(c)$17.07^{+3.05}_{-2.68}$&(c)$14.52^{+2.63}_{-2.30}$ &(c)$26.06^{+4.27}_{-3.72}$ \\ \hline
\hline
\end{tabular}\end{center}
\end{table}

In numerical calculations, we will use the following input
parameters (all masses are in GeV)
\beq m_d&=&5.4\times10^{-3},\quad m_s=0.15,\quad m_b=4.6,  \non
m_c&=&1.4, \quad \overline{m}_t(m_t)=165.9, \quad
m_{B_d}=5.279,\quad m_{B_s}=5.367, \non A&=&0.853, \quad
\lambda=0.225, \quad \bar{\rho}=0.20\pm0.09,\quad
\bar{\eta}=0.33\pm 0.05, \label{eq:input}
\eeq
where $A$, $\lambda$, $\bar{\rho}$ and $\bar{\eta}$ are Wolfenstein
parameters of the CKM mixing matrix.

For the hadronic parameters $f^2_{B_q} \hat{B}_{B_{q}}$, we use the values as given
by the JLQCD collaboration~\cite{jlqcd},
\begin{eqnarray}\label{eq:decay-constant}
f_{B_d} \hat{B}^{1/2}_{B_{d}}|_{\rm JLQCD} &=& (0.215 ^{+0.019}_{-0.030} ) \; {\rm GeV}, \nonumber\\
f_{B_s} \hat{B}^{1/2}_{B_{s}}|_{\rm JLQCD} &=& (0.245 ^{+0.021}_{-0.020} ) \;{\rm GeV},
\end{eqnarray}
where the individual errors given in Ref.~\cite{jlqcd} have been added in quadrature.

In a previous paper \cite{xiao06}, we studied the new physics contributions to the $B \to X_s \gamma$ decay,
and found strong constraints on the free parameters of the considered T2HDM:
\begin{enumerate}
\item[]{(i)}
A light charged Higgs boson with a mass less than $200$ GeV is excluded. For fixed $\tan{\beta}=30$ and
$\delta=0^\circ$, the lower limit on $M_H$ is $M_H \geq 300$ GeV.

\item[]{(ii)}
The data of $B \to X_s \gamma$ prefer a small angle $\delta$: $\delta < 44^\circ$ for
$\tan{\beta}=30$ and $M_H=400$ GeV.

\end{enumerate}
Here we will consider these constraints in our choice for the free parameters of the T2HDM.

The new physics (NP) contributions to $B^0 - \bar{B}^0$ mixing are
in general theoretically clean to interpret and have simple
operate structure. To constrain deviations from the SM in these
processes, we use the well measured physical observable $\Delta
M_d$  as well as the first observation of  $\Delta M_s$ as given in Eqs.(\ref{eq:exp1},\ref{eq:exp2})
and consider the effects of the hadronic uncertainty.
The theoretical predictions about the mass difference $\Delta M_q$ in the SM and
T2HDM are listed in Table~\ref{tab:deltam} for $\tan{\beta}=10,30,50$,
$M_H=300$ GeV and $\delta=0^\circ, 30^\circ$
and $60^\circ$. It is clear that the new physics contribution to
$B^0 - \ov{B}^0$ mixing is not sensitive to the parameter $\delta$ when $\tan \beta < 20$.
\begin{figure}[tbp]
\centerline{\mbox{\epsfxsize=8cm\epsffile{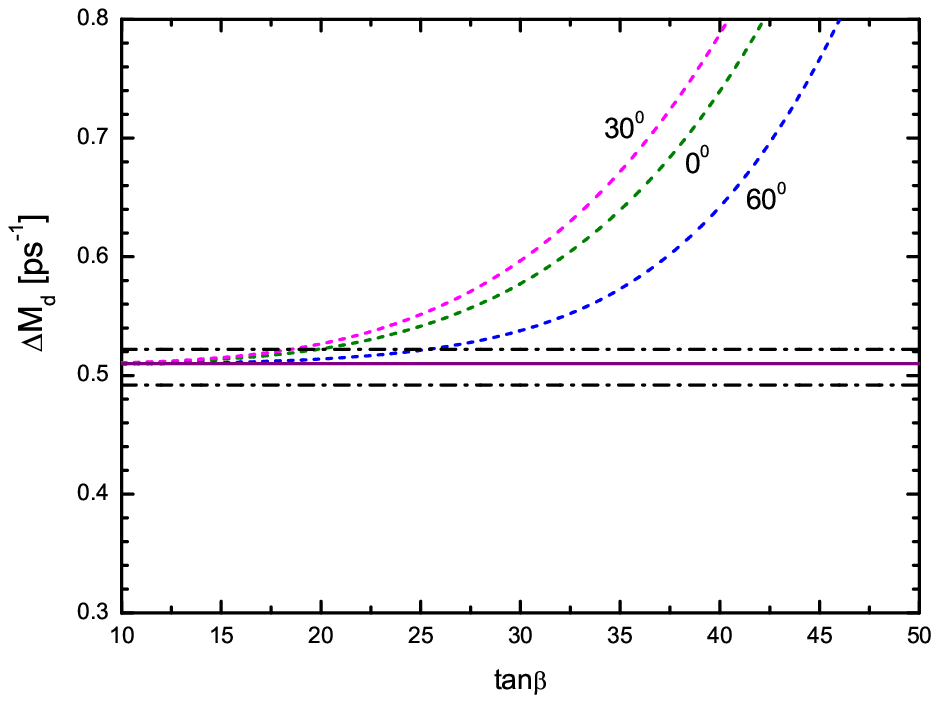}
\epsfxsize=8cm\epsffile{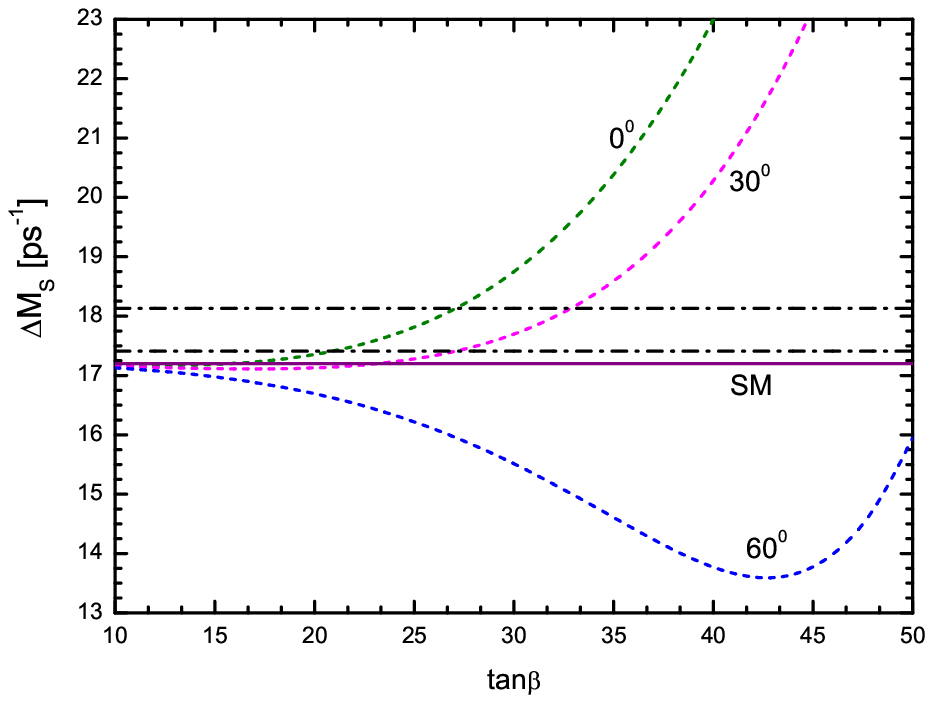}}}
\vspace{0.1cm}
\caption{The plots of $\Delta M_q$ vs $\tan{\beta}$ in the SM and T2HDM for $M_H=400$ GeV,
$\delta=0^\circ$, $30^\circ$ and $60^\circ$.}
\label{fig:fig1}
\end{figure}

In Fig.~\ref{fig:fig1}, we show the $\tan\beta$ dependence of $\Delta M_d$ and $\Delta M_s$
in the SM and T2HDM with the central values of hadronic parameters.
The band between two horizontal dot-dashed lines shows the measured values within $3\sigma$ errors:
$0.492 \leq \Delta M_d \leq 0.522$ (${\rm  ps^{-1}}$), and
$17.41 \leq \Delta M_s \leq 18.13$ (${\rm  ps^{-1}}$).
The solid horizontal lines show the central values of the SM predictions, which
agree well with the data. The dominant theoretical error comes from the large uncertainty of
hadronic parameter $f_{B_q} \hat{B}^{1/2}_{B_{q}}$.
The three short-dashed curves in Fig.~\ref{fig:fig1} represent the theoretical predictions
in the T2HDM for $M_H=400{\rm GeV}$ and for $\delta=0^\circ$, $30^\circ$ and $60^\circ$, respectively.
From this figure, the upper bound on $\tan\beta$ can be read off,
\be
\tan\beta \leq 25
\ee
for $M_H=400$ GeV and $\delta=[0^\circ,60^\circ]$. This bound is much stronger than the one
obtained from the radiative decay $B \to X_s \gamma$~\cite{xiao06}.
After the inclusion of the effects of the uncertainties of the hadronic parameters, the
upper bound on $\tan\beta$ will be changed into
\beq
\tan\beta \leq 30.
\eeq

\begin{figure}[tbp]
\centerline{\mbox{\epsfxsize=8cm\epsffile{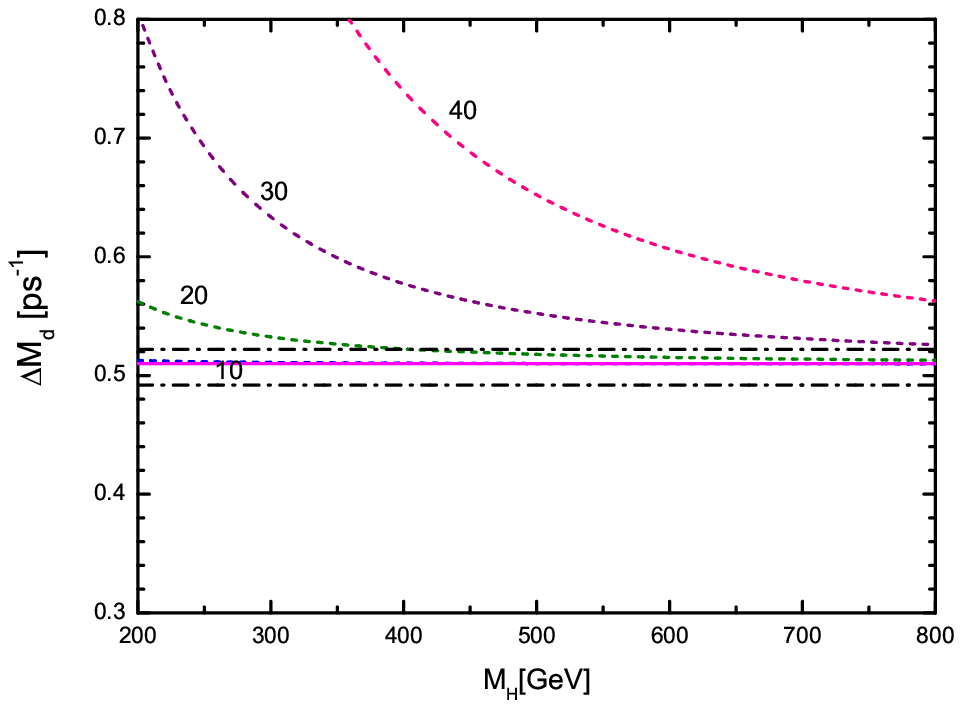}
\epsfxsize=8cm\epsffile{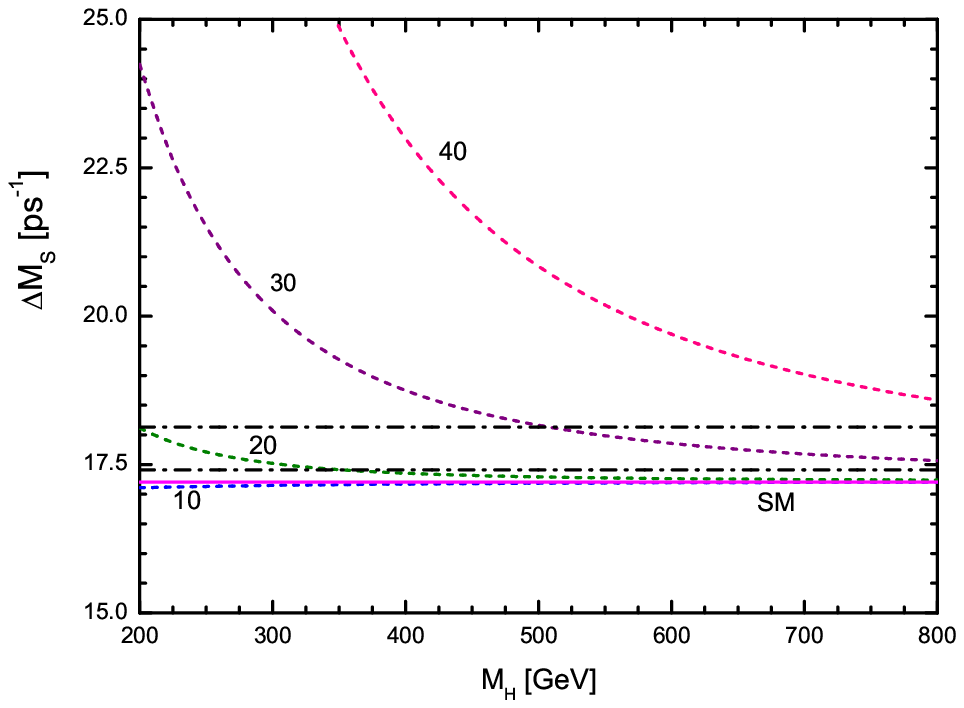}}}
\caption{The plots of $\Delta M_q$ vs $M_H$ in the SM and T2HDM, for $\delta=0^\circ$ and
for $\tan{\beta}=10$, $20$, $30$ and $40$, respectively. } \label{fig:fig2}
\end{figure}

In Fig.\ref{fig:fig2}, we show the $\tan\beta$ dependence of $\Delta
M_q$ in the SM and T2HDM for $\delta=0^\circ$ and
for $\tan{\beta}=10$, $20$, $30$ and $40$, respectively.
Here the central values of the input parameters are used.
The same as Fig.~\ref{fig:fig1}, the band between two horizontal dot-dashed lines shows
the measured values within $3\sigma$ errors.
The solid horizontal lines also show the central values of the SM predictions.
The four short-dashed curves are the theoretical predictions in the T2HDM for fixed
$\delta=0^\circ$, and for $\tan\beta=10$, $20$, $30$ and $40$, respectively.
It is easy to see that the data prefer a small $\tan{\beta}$, say $\tan{\beta} \leq 20$, if one
assumes the existence of a light charged Higgs boson with a mass around $300$ GeV.
The bounds on $\tan{\beta}$ and $M_H$ are indeed strongly correlated:
a smaller (larger) $\tan{\beta}$ means a lighter (heavier) charged Higgs boson.

%

\section{Conclusions}

In this paper, we have calculated the new physics contributions
to the neutral B meson mass splitting $\Delta M_d$ and $\Delta M_s$ induced by
the one-loop box diagrams involving one or two charged Higgs boson propagators
in the framework of T2HDM.

By comparing the theoretical predictions with the precision data of $B_{d(s)}^0-\bar{B}_{d(s)}^0$ mixing,
strong constraints on the free parameters of T2HDM can be obtained.
From the numerical results presented in last section, one can see that:
\begin{enumerate}
\item[]{(i)}
For fixed $M_H=400$ GeV and $\delta=[0^\circ,60^\circ]$, the upper bound on $\tan{\beta}$ is
\beq
\tan \beta \leq 30
\eeq
after the inclusion of major theoretical uncertainty.

\item[]{(ii)}
For a small $\tan{\beta}$, say $\tan{\beta} \leq 20$, a light charged Higgs boson with a mass around
$300$ GeV is allowed. The bounds on $\tan{\beta}$ and $M_H$ are indeed strongly correlated:
a smaller (larger) $\tan{\beta}$ means a lighter (heavier) charged Higgs boson.

\end{enumerate}

\begin{acknowledgments}
One of the authors Lin-xia L\"u would like to thank Professor C.-S.~Huang for his valuable help.
This work is partly supported  by the National Natural Science Foundation of China under Grant
No.10575052, and by the Specialized Research Fund for the doctoral Program of higher
education (SRFDP) under Grant No.~20050319008.

\end{acknowledgments}

\newpage

\end{document}